\documentclass[prd,aps,onecolumn,preprintnumbers,amsmath,amssymb,superscriptaddress]{revtex4}
\pdfoutput=1

\usepackage{amsmath}
\usepackage{graphicx}
\usepackage{dcolumn}
\usepackage{bm}
\usepackage{amssymb}
\usepackage{latexsym}
\usepackage{epstopdf}
\usepackage{threeparttable}
\usepackage{booktabs}
\usepackage{tabularx}
\usepackage{ulem}
\usepackage{float}
\usepackage{mathrsfs}
\usepackage{hhline}
\usepackage{multirow}
\usepackage{color}
\usepackage{subfigure}
\usepackage[colorlinks,linkcolor=magenta,anchorcolor=blue,citecolor=blue]{hyperref}

\makeatletter

\newcommand{\Rmnum}[1]{\expandafter\@slowromancap\romannumeral #1@}
\makeatother

\def\be{\begin{equation}}
	\def\ee{\end{equation}}
\def\bea{\begin{eqnarray}}
	\def\eea{\end{eqnarray}}

\begin{document}
\title{Can the Quantization of Black Hole be detected?}

\author{Ruifeng Zheng}
\email{zrf2021@mails.ccnu.edu.cn}
\affiliation{Institute of Astrophysics, Central China Normal University, Wuhan 430079, China}
	
\author{Taotao Qiu}
\email[Second author: ]{qiutt@hust.edu.cn}
\affiliation{School of Physics, Huazhong University of Science and Technology, Wuhan, 430074, China}

	\begin{abstract}
	As a mysterious celestial body predicted by General Relativity, black holes have been confirmed by observations in recent years. But there are still many unknown properties waiting for us to discover, one of the famous problems is the quantization of black holes. The quantization of black holes gives new research significance to black holes, but the quantization of black holes is difficult to prove. Here we propose a new method to verify the quantization of black holes: Carnot cycle. Our results show that if the black holes have quantized energy levels, the work $ W $ they impose to us via the Carnot cycle should also be quantized. We indirectly verify the quantization of black holes by detecting the quantization of $ W $, and contrast it with the classic Carnot cycle. In addition, we have also verified the correctness of the second law of thermodynamics under the premise of considering the quantization of black holes.  
	\end{abstract}
		\maketitle		
	\section{introduction}
  Since Bekenstein's pioneering work \cite{Bekenstein:1973ur}, black hole entropy has aroused great interest among people. In this famous paper, an important conception has been proposed: the minimum increase area of the black hole's horizon, which can furtherly lead to the quantization of black holes \cite{Bekenstein:1973ur, Bekenstein:1974jk}. In 1998, Hod derived a more accurate quantization condition of the Schwarzschild black hole surface area \cite{Hod:1998vk}: $ A_{n}=4\ln 3\cdot l_{p}^{2}\cdot n~ ~(n= 1,2,3\cdot \cdot \cdot n)$. Because the black hole surface area is proportional to the second power of the mass of the black hole, the quantization of the black hole surface area will also lead to the quantization of the mass of the black hole \cite{Corda:2019vuk, Hod:2015qfc, Sun:2018muq, He:2010ct}. Due to this, the energy absorbed or released by the black hole is no longer continuous \cite{Bekenstein:1995ju, Sun:2018muq, Bekenstein:1974jk}. Therefore, under the assumption of quantization, the mass of the black hole is related to the quantum number $ n $. In the following, we use the energy level of the black hole to describe the quantization of the mass of the black hole. 
  \par 
However, an interesting and important question is how to detect the energy level of the black hole. Due to the above reason, the discontinuous spectrum presents the characteristics of quantization, which provides us with the feasibility of detecting the energy level of the black hole. Here we propose another way: Carnot cycle.
  \par
  The Carnot cycle is a simple cycle in thermodynamics that only involves the transfer of energy. The simplest Carnot cycle consists of two heat sources, a high-temperature heat source, and a low-temperature heat source. Energy is transferred between the two heat sources through the working medium and does work. A complete Carnot cycle includes four steps: isothermal expansion, adiabatic expansion,  isothermal compression, and adiabatic compression.  At present, the Carnot cycle has been widely used in black hole physics \cite{Scandurra:2001pz, Ma:2019jmr, Zhang:2016wek, Wei:2016hkm, Hendi:2017bys, Curiel:2014zua, Mo:2017nes, Jacobson:2018nnf, Kaburaki:1991zz, Wei:2017vqs, Guo:2021bju, Johnson:2014yja, Mo:2017nhw, Ghaffarnejad:2018gtj, Prunkl:2019wdw, Feng:2021vey}.
  \par
  Since the Carnot heat engine inevitably does work, the efficiency of the Carnot cycle can be expressed as $ \eta =W/Q $, where $ W $ is the work, and $ Q $ is the energy absorbed by the working medium from the high-temperature heat source. It can be seen from the efficiency expression of the Carnot cycle that for a constant $ Q $, the greater the work $ W $, the higher the efficiency. Similarly, we can also use temperature to express the efficiency of the Carnot cycle \cite{Scandurra:2001pz}: $ \eta =1-{T_
 l}/{T_h} $, where $ T_{h} $ is the temperature of the high-temperature heat source, and $ T_{l} $ is the temperature of the low-temperature heat source. 
  \par
  Since the Carnot cycle of black holes only involves the transfer of energy, we naturally associate it with the energy level hypothesis of black holes. The energy level hypothesis of black holes indicates that the energy absorbed or released by a black hole is equal to the energy level interval of the black hole. Therefore, the value of the energy transfer of the Carnot cycle of the black hole should be related to the difference of energy level of the energy level of the black hole, and the corresponding Carnot cycle efficiency as well as the work $ W $ should also be related to the quantum number of the black hole. By determining whether the work $ W $ is quantized, the Carnot cycle of the black hole provides us with a reasonable method to detect the existence of black hole energy level. 
  \par 
 The rest of our paper is organized as follows: In Sec. \ref{sec2}, starting from the quantization condition of the surface area of the Schwarzschild black hole, we obtain some quantization relations about the Schwarzschild black hole. In Sec. \ref{sec3}, we introduce the four processes of the Carnot cycle, small/large black holes are used as high-temperature/low-temperature heat sources, respectively. In Sec. \ref{sec4}, we give the expression of the efficiency $ \eta_{oc} $ of the Carnot cycle under the assumption of quantization, and show that the work $ W_{oc}$ can also be quantized. We further discuss the evolution of entropy in the Carnot cycle. In Sec. \ref{sec5}, we come to our conclusions and discussions.  
  	\section{Quantization of black holes}
  	\label{sec2}
  	Bekenstein pointed out in his famous paper that the limit of a point particle is illegal in quantum theory \cite{Bekenstein:1973ur, Hod:1998vk}. This violates the Heisenberg uncertainty relationship. Bekenstein argued that particles have a finite proper radius $ b $, and showed that the assimilation of a finite size neutral particle inevitably causes an increase in the surface area:
  		\begin{equation}\label{EQ01}
  	\left( \varDelta A \right) _{\min}=8\pi \mu b~,
  	 	\end{equation}
   	where $ A $ is the black hole surface area and $ \mu $ is the rest mass of the particle. But we know that the size of a relativistic particle cannot be smaller than its Compton wavelength: $ b\ge {h}/{4\pi \mu c} $. This yields a lower bound on the increase in the black hole surface area due to the assimilation of a (neutral) test particle \cite{Bekenstein:1973ur}:
   		\begin{equation}\label{EQbk}
   	\left( \varDelta A \right) _{\min}=8\pi l_{p}^{2}~~~~(neutral~~particle),
   		\end{equation}
   	where $ l_{p} $ is the Planck length. While Eq.(\ref{EQ01}) only applies to the neutral particle, Hod found that the lower bound of the area increase caused by the absorption of the charged particle is \cite{Hod:1999oyc}
   		\begin{equation}\label{EQHod}
   	\left( \varDelta A \right) _{\min}=4l_{p}^{2}~~~~(charged~~particle).
   		\end{equation}
  Therefore, the quantization condition of the black hole surface area can be expressed as \cite{Hod:1998vk}    
			\begin{equation}\label{EQ1}
	A_n=\alpha l_{p}^{2}\cdot n\ \ \left( n=1,2,3\cdot \cdot \cdot \right)~.
		\end{equation}
For the Schwarzschild black hole, $ \alpha =4\ln 3 $ \cite{Hod:1998vk}. On the other hand, the black hole surface area is given by 
		\begin{equation}\label{EQ2}
	A=4\pi r_{h}^{2}~,
		\end{equation}
where $ r_{h} $ is the black hole horizon radius. Moreover, the relationship between the mass $ M $ and the radius $ r_{h} $ of the Schwarzschild black hole is 
		\begin{equation}\label{EQ5}
	r_{h}=\frac{2GM}{c^2}~.
		\end{equation}
	  	\par
Combining Eq.(\ref{EQ2}) and Eq.(\ref{EQ5}), we get 
		\begin{equation}\label{EQM}	
M=\frac{c^2}{4G}\sqrt{\frac{A}{\pi}}~.
	    \end{equation}
Substituting Eq.(\ref{EQ1}) into Eq.(\ref{EQM}), the quantized mass of the Schwarzschild black hole can be written as
			\begin{equation}\label{EQ6}		
M_n=\frac{m_p}{4}\sqrt{\frac{\alpha}{\pi}}\sqrt{n}~,		
			\end{equation}
where $m_p$ is the Planck mass. Therefore, the energy level of the Schwarzschild black hole can be written as 		
	\begin{equation}\label{EQ7}	
E_n=M_{n}c^2=\frac{m_{p}c^2}{4}\sqrt{\frac{\alpha}{\pi}}\sqrt{n}~,
			\end{equation}
and the corresponding energy level interval can be written as ($ n>m $)
		\begin{equation}\label{EQ8}	
	\varDelta E_{m\rightarrow n}=E_{n}-E_{m}=\frac{m_{p}c^2}{4}\sqrt{\frac{\alpha}{\pi}}\left( \sqrt{n}-\sqrt{m} \right)~.
	\end{equation}	
\begin{figure}[htbp]
	\centering
	\includegraphics[height=6.0cm,width=9.75cm]{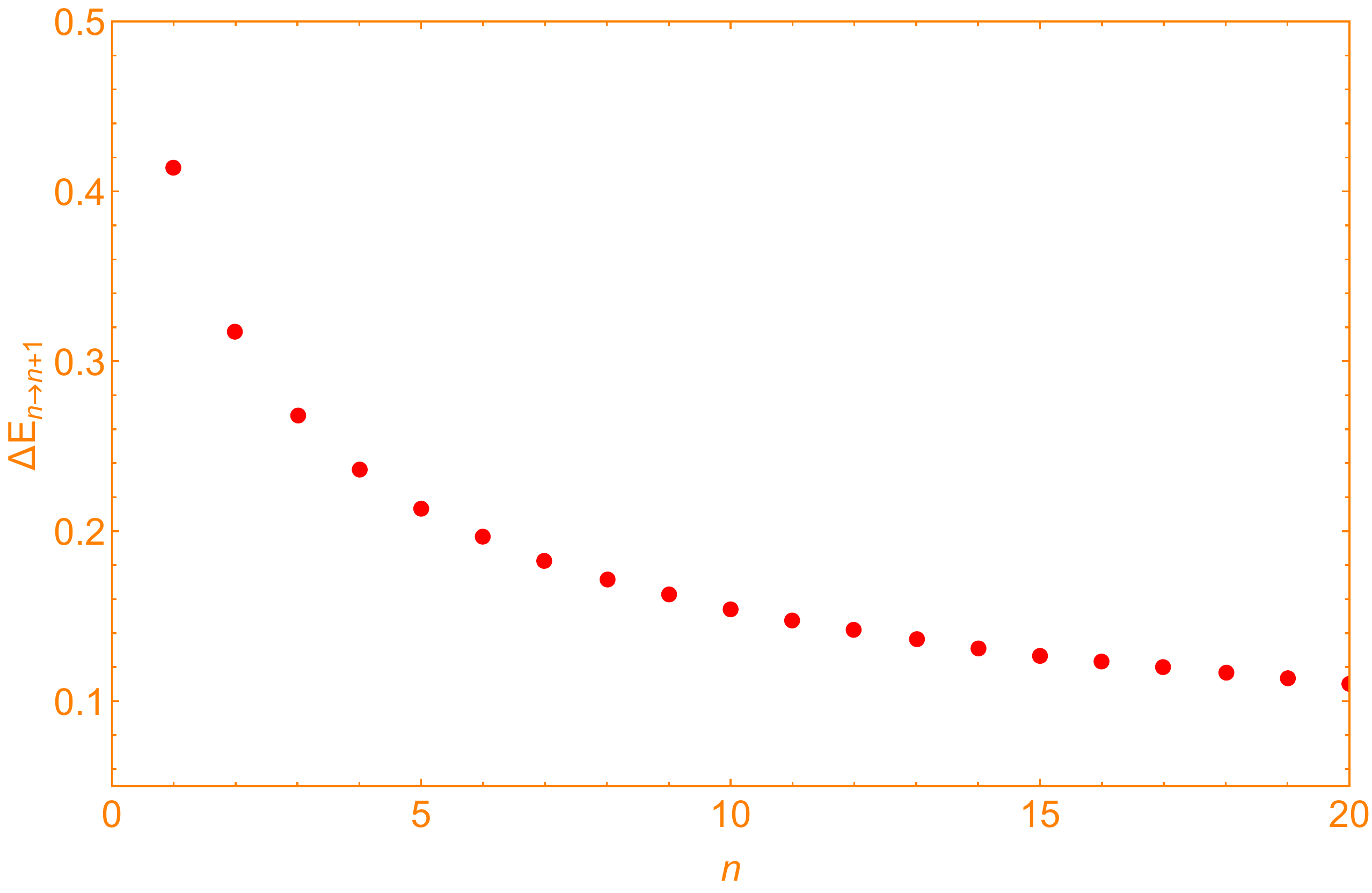}
	\caption{The red dot represents the quantization of the energy level interval of the black hole, we ignore the correlation coefficient. }
	\label{FIG1}
\end{figure}
It can be seen from Eq.(\ref{EQ8}) that as the quantum number $ n $ of black holes increases, the energy level interval $ \varDelta E_{n\rightarrow n+1} $ of the black hole becomes smaller and smaller. We plot Fig. \ref{FIG1} to show this relation.
\par
\begin{figure}[htbp]\label{FIG2}
		\includegraphics[height=6.0cm,width=9.75cm]{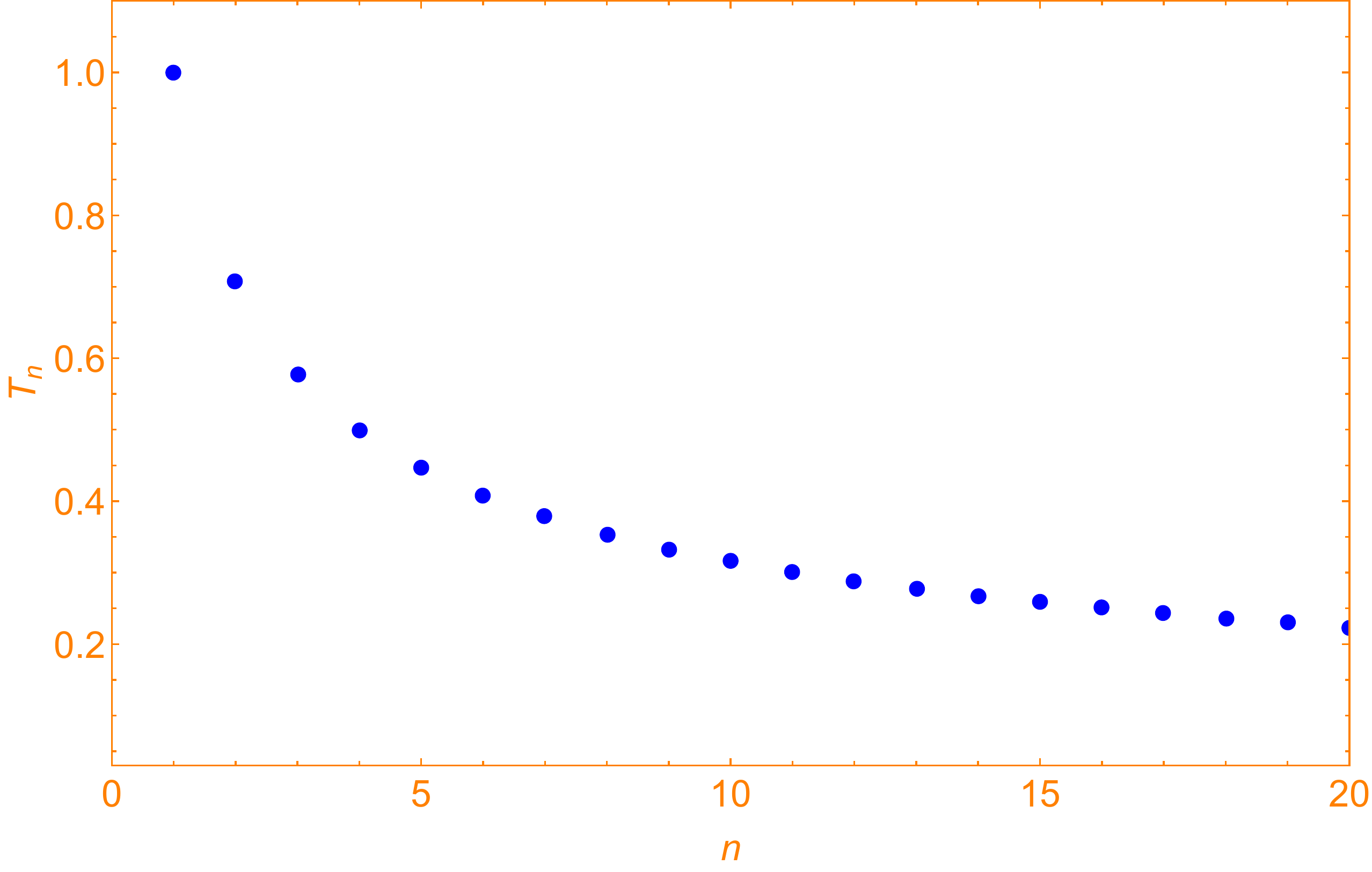}
		\caption{The blue dot represents the quantization of the black hole temperature, we ignore the correlation coefficient. }
	\end{figure}

	
Similarly, the expression of the horizon temperature $ T $ and mass $ M $ of  the Schwarzschild black hole is 
			\begin{equation}\label{EQ9}
		T=\frac{hc^{3}}{16\pi ^2kGM}~.
			\end{equation}
Substituting Eq.(\ref{EQ6}) into Eq.(\ref{EQ9}), we get \cite{He:2010ct}
		\begin{equation}\label{EQ10}	
	T_n=\frac{hc^{3}}{4\pi ^2kGm_p}\sqrt{\frac{\pi}{\alpha}}\frac{1}{\sqrt{n}}~.	
		\end{equation}
Setting $ \beta =\frac{hc^{3}}{4\pi ^2kGm_p}\sqrt{\frac{\pi}{\alpha}} $, Eq.(\ref{EQ10}) is simplified to 
		\begin{equation}
	T_n=\beta \frac{1}{\sqrt{n}}~.
		\end{equation}
Therefore, we get the relationship between the horizon temperature $ T_{n} $ of the Schwarzschild black hole and the quantum number $ n $. We plot Fig. \ref{FIG2} to show this relation. Furthermore, we can also speculate that the radiation spectrum of the black hole should also be related to the quantum number $ n $, however, since it is beyond the scope of this paper, we will postpone it for the further discussion. 

	\section{Carnot cycle}
		\label{sec3}	
\begin{figure}[htbp]\label{FIG3}
        \centering
		\includegraphics[height=8.0cm,width=14cm]{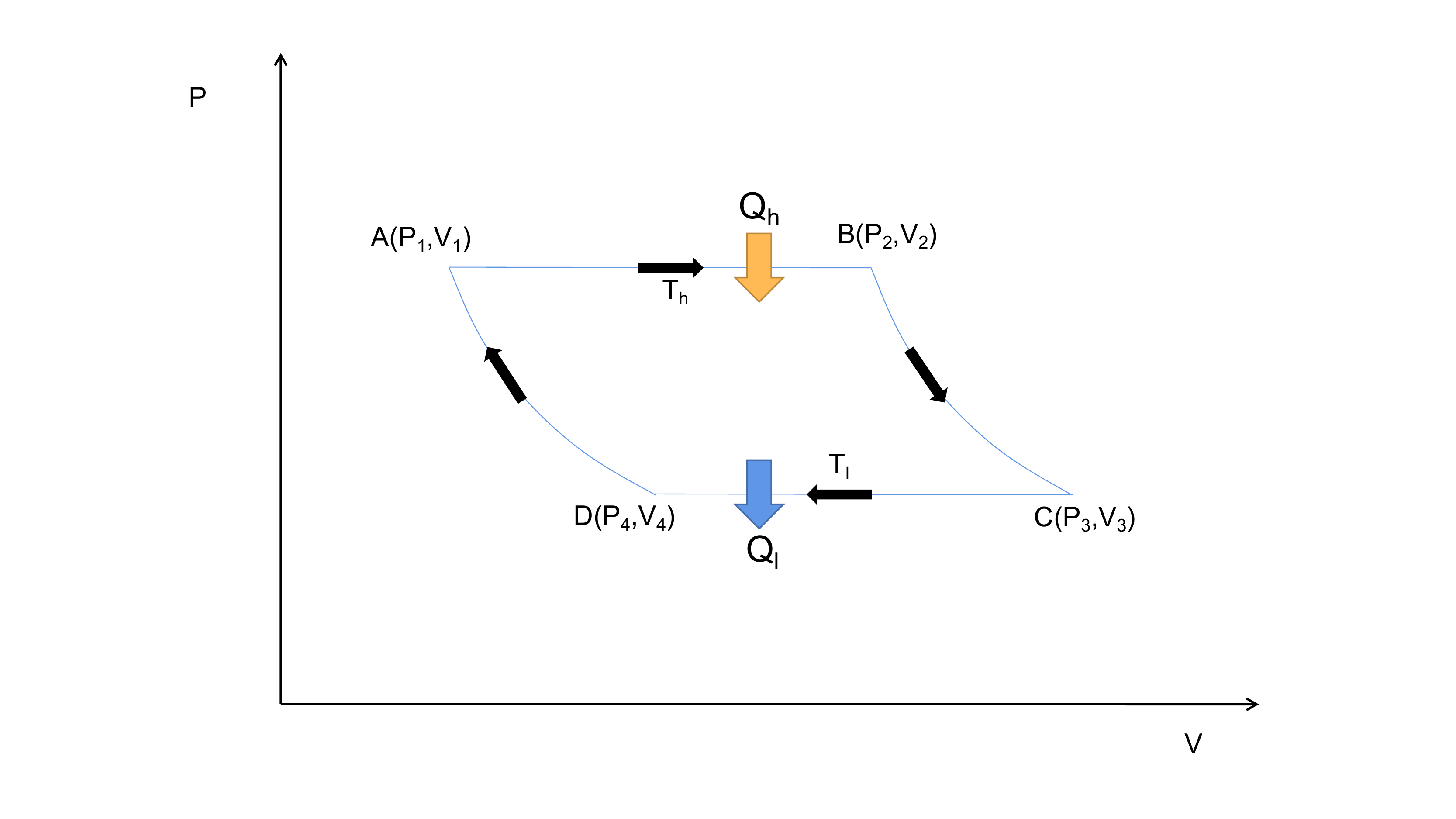}
		\caption{We use the radiation gas as the working matter, A$\rightarrow$B, B$\rightarrow$C, C$\rightarrow$D and D$\rightarrow$A to represent the isothermal expansion, adiabatic compression, isothermal compression, and adiabatic expansion respectively. $T_h$ represents the temperature of the high-temperature heat source, $T_l$ represents the temperature of the low-temperature heat source. $Q_h$ is the energy absorbed by the high-temperature heat source (small black hole), and $Q_l$ is the energy released to the low-temperature heat source (large black hole). Since A$\rightarrow$B and C$\rightarrow$D are isothermal processes, we have $P_1=P_2=aT_{h}^{4}/3$, $P_3=P_4=aT_{l}^{4}/3$, where $a=\pi ^2k^4/15c^3{\hbar }^3$.}
	\end{figure}

In this part, we briefly introduce the four processes of the Carnot cycle, and draw a P-V diagram Fig. \ref{FIG3} with radiation gas as an example.
\par
The fist step is isothermal expansion. In this step, the working medium absorbs energy $Q_h$ from a high-temperature $T_{h}$ heat source and expands isothermally from point A($P_1,V_1$) to point B($P_2,V_2$). So we have
\begin{equation}
   \varDelta U_1=0~, 
\end{equation}
\begin{equation}
W_1=nRT_h\ln \frac{V_2}{V_1}~,
\end{equation}\\
where $n$ is the amount of substance in $mol$, $R$ is the molar gas constant. According to the first law of thermodynamics, we have
\begin{equation}
Q_{h}=W_1~.
\end{equation}
\par
The second step is adiabatic expansion, the working medium expands adiabatically from point B($P_2,V_2,T_h$) to point C($P_3,V_3,T_l$). In this step:
\begin{equation}
Q_{2}=0~,
\end{equation}
\begin{equation}
W_2=-\bigtriangleup  U_2=-\int_{T_h}^{T_l}{C_{V,m}dT}~,
\end{equation}
where $C_{V,m}$ is the molar heat capacity at constant volume.
\par
The thrid step is isothermal compression, the working medium isothermal compression from point C($P_3,V_3$) to point D($P_4,V_4$). In this step:
\begin{equation}
   \varDelta U_3=0~, 
\end{equation}
\begin{equation}
W_3=nRT_l\ln \frac{V_4}{V_3}~.
\end{equation}\\
According to the first law of thermodynamics, we have
\begin{equation}
Q_{l}=W_3~.
\end{equation}
\par
The fourth step is adiabatic compression, the working medium adiabatic compression from point D($P_4,V_4,T_l$) to point A($P_1,V_1,T_h$). In this step, we have
\begin{equation}
Q_{4}=0~,
\end{equation}
\begin{equation}
W_4=-\bigtriangleup  U_4=\int_{T_h}^{T_l}{C_{V,m}dT}~.
\end{equation}
\par
So in the whole cycle
\begin{equation}
\bigtriangleup U=0~,
\end{equation}
\begin{equation}
Q=Q_h+Q_l~,
\end{equation}
\begin{equation}
W=W_1+W_3~.
\end{equation}
According to the adiabatic reversible process equation, in the second process:
\begin{equation}
T_hV_{2}^{\gamma -1}=T_lV_{3}^{\gamma -1}~,
\end{equation}
in the fourth process:
\begin{equation}
T_hV_{1}^{\gamma -1}=T_lV_{4}^{\gamma -1}~,
\end{equation}
then we have
\begin{equation}
\frac{V_2}{V_1}=\frac{V_3}{V_4}~.
\end{equation}
Therefore, during the whole cycle, the work
\begin{equation}
W=W_1+W_3=nRT_h\ln \frac{V_2}{V_1}+nRT_l\ln \frac{V_4}{V_3}=nR\left( T_h-T_l \right) \ln \frac{V_2}{V_1}
\end{equation}
The cycle efficiency is
\begin{equation}\label{EQ.31}
\eta =\frac{W}{Q_h}=\frac{nR\left( T_h-T_l \right) \ln \frac{V_2}{V_1}}{nRT_h\ln \frac{V_2}{V_1}}=\frac{T_h-T_l}{T_h}=1-\frac{T_l}{T_h}
\end{equation}
\par 
	\section{Quantization of the Real Cycle }
	\label{sec4}
In this part, we use small black hole and large black hole to represent high-temperature $T_h$ heat sources and low-temperature $T_l$ heat sources, respectively. The Carnot cycle shows that the small black hole releases energy as the high-temperature heat source, while the large black hole absorbs energy as the low-temperature heat source. In the second part of this paper, the quantization of the black hole shows that the black hole absorbs or releases energy corresponding to the energy level interval. Therefore, we think that the efficiency $ \eta $ of the Carnot cycle should be related to the quantum number $ n $ of the black hole, $ Q_{h} $, $ Q_{l} $ corresponds to the energy level interval of the black hole: 
\begin{equation}\label{EQ26}
Q_{h}=\left( M_m-M_{m-\varDelta m} \right)c^{2}~~(Small~black~hole~transition)~, 
\end{equation}
\begin{equation}\label{EQ27}
Q_{l}=\left( M_{n+\varDelta n}-M_n \right)c^{2}~~(Large~black~hole~transition)~, 
\end{equation}
where we use $ n $ to represent the quantum number of the large black hole, $ m $ to represent the quantum number of the small black hole, $ \varDelta n $ to indicate the energy level of the transition of the large black hole, and $ \varDelta m $ to indicate the energy level of the transition of the small black hole.
It can be easily concluded that the work of a period of the Carnot cycle is 
\begin{equation}\label{EQ28}
W_{oc}=Q_{h}-Q_{l}=(M_n+M_m-M_{n+\varDelta n}-M_{m-\varDelta m})c^{2}~,
\end{equation}
Therefore, we can get the work is also quantized: 
\begin{equation}\label{EQ32}
W_{oc}=\frac{m_{p}c^{2}}{4}\sqrt{\frac{\alpha}{\pi}}\left( \sqrt{n}+\sqrt{m}-\sqrt{n+\varDelta n}-\sqrt{m-\varDelta m} \right)~. 
\end{equation}
We substitute Eq.(\ref{EQ26}) and Eq.(\ref{EQ28}) into Eq.(\ref{EQ.31}) to get
\begin{equation}\label{EQ29}
\eta_{oc} =1-\frac{M_{n+\varDelta n}-M_n}{M_m-M_{m-\varDelta m}} =1-\frac{\sqrt{n+\varDelta n}-\sqrt{n}}{\sqrt{m}-\sqrt{m-\varDelta m}}~,
\end{equation}
	\begin{figure}[htbp]
	\centering
	\subfigure{
		\includegraphics[height=4.0cm,width=7.5cm]{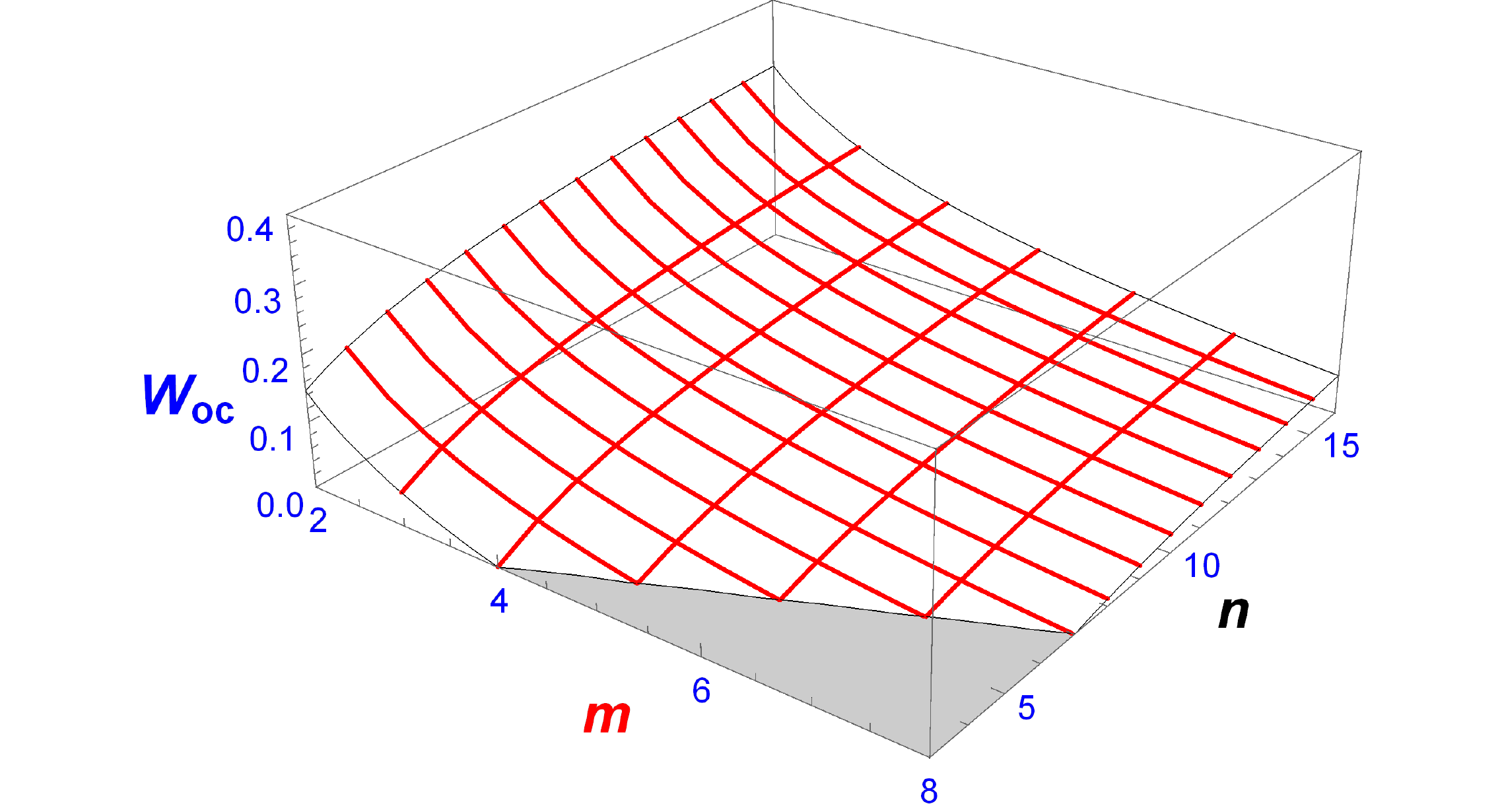}
	}
	\subfigure{
		\includegraphics[height=4.0cm,width=6.5cm]{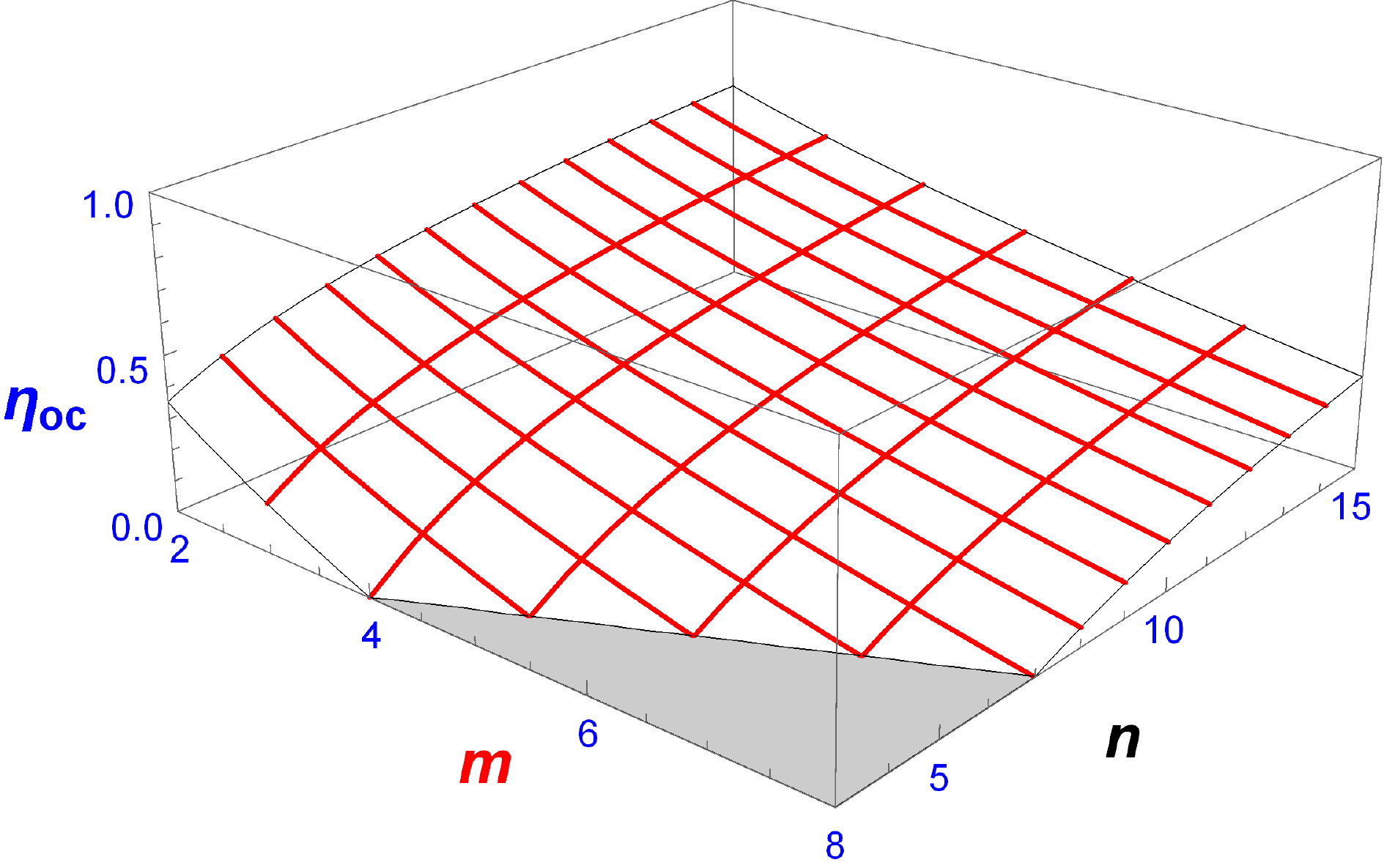}
	}
	\caption{{\it The left figure:} $ W_{oc} $ in quantized form, the intersection of the network lines in the figure corresponds to the value in Eq. (\ref{EQ32}). {\it The right figure:} $ \eta_{oc} $ in quantized form, the intersection of the network lines in the figure corresponds to the value in Eq. (\ref{EQ29}). Where $ n $ is the quantum number of the large black hole, $ m $ is the quantum number of the small black hole: $ n > m $, we set $ \varDelta n=\varDelta m=1 $. Because the efficiency of the 
cycle cannot be less than 0, the shaded area is excluded.  }
	\label{FIG5}
\end{figure}
where $ \varDelta m $ and $ \varDelta n $ take positive integers. As an example, we plot Fig. \ref{FIG5} with $ \varDelta n=\varDelta m=1 $. 
\par
We know that the efficiency of the Carnot cycle is related to the temperature of the heat source, combining Eq.(\ref{EQ10}) and Eq.(\ref{EQ.31}), we get 
\begin{equation}\label{EQ30}
\eta_{tv} =1-\frac{T_l}{T_h}=1-\sqrt{\frac{m-\varDelta m}{n+\varDelta n}}~,
\end{equation}
the subscript $ tv $ is the cycle efficiency value expressed in temperature, which is related to the quantum number of the black hole. But it should be noted that the real cycle efficiency  $ \eta_{oc} $ is always lower than the reversible cycle efficiency $ \eta_{tv} $:
\begin{equation}\label{EQ31}
\sqrt{\frac{m-\varDelta m}{n+\varDelta n}}< \frac{\sqrt{n+\varDelta n}-\sqrt{n}}{\sqrt{m}-\sqrt{m-\varDelta m}}~.
\end{equation}
\par
Since the horizon temperature of the black hole will change with the mass, as the Carnot cycle proceeds, the temperature difference between the two black holes will change, which causes the efficiency of the Carnot cycles different cycle by cycle. This leads to the quantization of the Carnot cycle efficiency, which depends on $ \varDelta n $ and $ \varDelta m $. For example, we set $ \varDelta n=\varDelta m=1 $ (each cycle transitions only one energy level). But this is not absolute.

\par
We take 4 sets of data and set $ \varDelta n,\varDelta m=1 $ to calculate $ \eta_{oc} $ and $ \eta_{tv} $. The results are shown in Table. \ref{Table1}. It can be seen from Table. \ref{Table1} that the value of $ \eta_{oc} $ is smaller than $ \eta_{tv} $, which is reasonable. 
\begin{table}
	\centering\caption{$ n $ is the quantum number of the large black hole, $ m $ is the quantum number of the small black hole, we set $ \varDelta n,\varDelta m=1 $, $ \eta _{tv} $ is the cycle efficiency value expressed in temperature, and $ \eta _{oc} $ is the real cycle efficiency. }
	\par
	\begin{tabular}{|c|c|c|c|c|c|} 
		\hline\rule{0pt}{10pt}
		$ m $ & $ n $	& $ \varDelta n $  &  $ \varDelta m $ &  $ \eta_{tv} $&  $ \eta_{oc} $ \\
		\hline\rule{0pt}{10pt}
		670 &	680 & 1 & 1 & 0.0088 & 0.0081 \\
		\hline\rule{0pt}{10pt}
		2300 & 2340	& 1 & 1 & 0.009 & 0.0088  \\
		\hline\rule{0pt}{10pt}
		1500 & 1520	& 1 & 1 & 0.0073 & 0.0069 \\
		\hline\rule{0pt}{10pt}
		4200 & 4250 & 1 & 1 & 0.0061 & 0.0060 \\
		\hline
	\end{tabular}\\
	\label{Table1}
\end{table}
\par
Please note that our choice of $ \varDelta n $, $ \varDelta m $=1 is not strict. It can be seen from Fig. \ref{FIG1} that the energy level interval of the large black hole is smaller than the energy level interval of the small black hole, which leads to a conclusion: two black holes with significantly different quantum numbers, the energy released by the jump of the small black hole to one energy level are enough to make the large black hole jump to several energy levels. There are several possible values for the real cycle efficiency $ \eta _{oc} $. But if the quantum number of the two black holes are not significantly different, the energy released by the small black hole jumping to one energy level can only cause the large black hole to jump to one energy level. 
\par 
We further explore the evolution of black hole entropy in the cycle. The black hole entropy is proportional to the black hole surface area $ S_{BH}\propto A $ \cite{Bekenstein:1973ur}, and the surface area is proportional to the square of the mass of the black hole $ A\propto M^2 $, so we have 
	\begin{equation}\label{EQ11}
	S_{BH}=4\pi k\left( \frac{M}{m_p} \right) ^2~,
\end{equation}
where $k$ is the Boltzmann constant. From Eq.(\ref{EQ6}) and Eq.(\ref{EQ11}), the entropy of the black hole is proportional to the quantum number $ n $ \cite{Corda:2014wga, Cristofano:2013tla, Sakalli:2011zz, Corichi:2006wn}: 
\begin{equation}\label{EQ33}
S_{BH}\propto n~.
\end{equation}
\par
From the above analysis, it can be seen that the energy released by the small black hole jumping to one energy level is enough to make the large black hole jumping to at least one energy level, which means that $ \varDelta m=1 $, $ \varDelta n  \ge 1 $.
Therefore, the initial entropy of two black holes can be written as (we omit the coefficient)
\begin{equation}\label{EQ34}
S_{initial}\propto n+m~.
\end{equation}
The final entropy of the two black holes after one cycle is 
\begin{equation}\label{EQ35}
S_{final} \propto n+m-\varDelta m+\varDelta n~.
\end{equation}
Since $ \varDelta n \ge \varDelta m $, we have $ S_{final} \ge S_{initial} $, which shows that the total entropy of the system will never decrease during the evolution process, which also confirms the second law of thermodynamics.  
  
	\section{Conclusion}
	\label{sec5}
Through the quantization of the Schwarzschild black hole surface area, we got the quantization of the mass of the black hole. The latter indicated that the black hole can only release or absorb certain amount energy. In this paper, we proposed a possible way to detect the quantization hypothesis of black hole, namely Carnot cycle. 
\par	
The research on the Carnot cycle of the black hole has been widely discussed in the literature. In this paper, we adopted the classic Carnot cycle process (isothermal expansion, adiabatic expansion, isothermal compression, and adiabatic compression), and associated it with the quantization of black holes. The results shew that due to the quantization of the mass of the black hole, the energy transfer of the Carnot cycle of the black hole also presented the characteristics of quantization. We gave the corresponding cycle efficiency $ \eta_{oc} $, indicating that it is related to the quantum number $ n $, $ m $, $ \varDelta n $, $ \varDelta m $. Similarly, we also gave the quantized form of $ W_{oc} $.
\par 
Although it is difficult for us to directly detect the energy level of the black hole, the work $ W_{oc} $ in the cycle provided us with a way to observe whether the black hole has an energy level. By detecting the quantization of $ W_{oc} $, it could indirectly prove the quantization of the black hole, which is of great significance for us to deeply understand the microstructure of the black hole.
\par 
We further verified the correctness of the second law of thermodynamics under the assumption of black hole quantization. Since the energy released by the small black hole transitioning to one energy level is enough to make the large black hole transitioning to at least one energy level, the final entropy of the double black holes after one cycle would not be less than the initial entropy of the double black holes before the cycle: $ S_{final} \ge S_{initial} $.
\par
Our work provided a possibility to indirectly verify the energy level of the black hole. However, how to realize the Carnot cycle of double black holes in reality, is still a difficult problem. We expect the development of science and technology will give rise to progress along this way in the future.
 
\begin{acknowledgements}
We thanks Shao-Wen Wei for useful discussions and a careful reading of this
manuscript. T.Q. is supported in part by the National Natural Science Foundation of China under Grants No.~11875141, and in part by the National Key R\&D Program of China (2021YFC2203100).
\end{acknowledgements}

	\bibliographystyle{apsrev4-1}
	\bibliography{kanuo_1.bib}
\end{document}